\documentclass[a4paper,10pt, french]{article}
\usepackage[utf8x]{inputenc}
\usepackage[standard]{ntheorem}
\usepackage[francais]{babel}

\usepackage{amsmath}
\usepackage{color}
\usepackage{dsfont}

\title{A Complexity Approach for Steganalysis}
\author{Jacques M. Bahi, Christophe Guyeux, and Pierre-Cyrille Heam~\thanks{Authors in alphabetic order}}

\begin{document}

\maketitle

\begin{abstract}
In this proposal for the Journ\`ees Codes et St\'eganographie 2012, we define
a new rigorous approach for steganalysis based on the complexity theory. It is 
similar to the definitions of security that can be found for hash functions,
PRNG, and so on. We propose here a notion of \emph{secure hiding}
and we give a first secure hiding scheme. 
\end{abstract}

\section{Introduction}

Robustness and security are two major concerns in information hiding.
These two concerns have been defined in \cite{Kalker2001} as follows. 
``Robust watermarking is a mechanism to create a communication channel that is multiplexed into original content [...]. It is required that, firstly, the perceptual degradation of the marked content [...] is minimal and, secondly, that the capacity of the watermark channel degrades as a smooth function of the degradation of the marked content. [...]. Watermarking security refers to the inability by unauthorized users to have access to the raw watermarking channel. [...] to remove, detect and estimate, write or modify the raw watermarking bits.''

In the framework of watermarking and steganography, security has seen several important developments since the last decade~\cite{BarniBF03,Cayre2005,Ker06}. 
The first fundamental work in security was made by Cachin in the context of steganography~\cite{Cachin2004}. 
Cachin interprets the attempts of an attacker to distinguish between an innocent image and a stego-content as a hypothesis testing problem. 
In this document, the basic properties of a stegosystem are defined using the notions of entropy, mutual information, and relative entropy. 
Mittelholzer, inspired by the work of Cachin, proposed the first theoretical framework for analyzing the security of a watermarking scheme~\cite{Mittelholzer99}.

These efforts to bring a theoretical framework for security in steganography and watermarking have been followed up by Kalker, who tries to clarify the concepts (robustness \emph{vs.} security), and the classifications of watermarking attacks~\cite{Kalker2001}. 
This work has been deepened by Furon \emph{et al.}, who have translated Kerckhoffs' principle (Alice and Bob shall only rely on some previously shared secret for privacy), from cryptography to data hiding~\cite{Furon2002}. 
They used Diffie and Hellman methodology, and Shannon's cryptographic framework~\cite{Shannon49}, to classify the watermarking attacks into categories, according to the type of information Eve has access to~\cite{Cayre2005,Perez06}, namely: Watermarked Only Attack (WOA), Known Message Attack (KMA), Known Original Attack (KOA), and Constant-Message Attack (CMA).
Levels of security have been recently defined in these setups.
The highest level of security in WOA is called stego-security \cite{Cayre2008}, recalled below.

In the prisoner problem of Simmons~\cite{Simmons83}, Alice and Bob are in jail, and they want to, possibly, devise an escape plan by exchanging hidden messages in innocent-looking cover contents. 
These messages are to be conveyed to one another by a common warden, Eve, who over-drops all contents and can choose to interrupt the communication if they appear to be stego-contents. 
The stego-security, defined in this
framework, is the highest security level in WOA setup~\cite{Cayre2008}.
To recall it, we need the following notations:
\begin{itemize}
  \item $\mathds{K}$ is the set of embedding keys,
  \item $p(X)$ is the probabilistic model of $N_0$ initial host contents,
  \item $p(Y|K_1)$ is the probabilistic model of $N_0$ watermarked contents.
\end{itemize}

Furthermore, it is supposed in this context that each host content has been watermarked with the same secret key $K_1$ and the same embedding function $e$.
It is now possible to define the notion of stego-security:

\begin{definition}[Stego-Security]
\label{Def:Stego-security}
The embedding function $e$ is \emph{stego-secure} if and only if:
$$\forall K_1 \in \mathds{K}, p(Y|K_1)=p(X).$$
\end{definition}

\section{Toward a Cryptographically Secure Hiding}

\subsection{Introduction}

Almost all branches in cryptology have a complexity approach for security.
For instance, in a cryptographic context, a pseudorandom number generator (PRNG) is a deterministic
algorithm $G$ transforming strings  into strings and such that, for any
seed $k$ of length $k$, $G(k)$ (the output of $G$ on the input $k$) has size
$\ell_G(k)$ with $\ell_G(k)>k$.
The notion of {\it secure} PRNGs can now be defined as follows. 

\begin{definition}
A cryptographic PRNG $G$ is secure if for any probabilistic polynomial time
algorithm $D$, for any positive polynomial $p$, and for all sufficiently
large $k$'s,
$$| \mathrm{Pr}[D(G(U_k))=1]-Pr[D(U_{\ell_G(k)})=1]|< \frac{1}{p(k)},$$
where $U_r$ is the uniform distribution over $\{0,1\}^r$ and the
probabilities are taken over $U_N$, $U_{\ell_G(N)}$ as well as over the
internal coin tosses of $D$. 
\end{definition}

Intuitively, it means that no polynomial-time algorithm can make a 
distinction, with a non-negligible probability, between a truly 
random generator and $\mathcal{G}$.

Inspired by these kind of definitions, we propose what follows.

\subsection{Definition of a stegosystem}

\begin{definition}[Stegosystem]
Let $\mathfrak{A}$ an alphabet and $\mathcal{S}, \mathcal{M}, \mathcal{K}$ 
three sets of words  on $\mathfrak{A}$ called respectively the sets of 
supports, messages, and keys.

A \emph{stegosystem} on $(\mathcal{S}, \mathcal{M}, \mathcal{K})$  
is a tuple $(\mathcal{I},\mathcal{E}, inv)$ such that:
\begin{itemize}
\item $\mathcal{I}:\mathcal{S} \times \mathcal{M} \times \mathcal{K} \longrightarrow \mathcal{S}$, 
$(s,m,k) \longmapsto \mathcal{I}(s,m,k)=s'$, 
\item $\mathcal{E}:\mathcal{S}  \times \mathcal{K} \longrightarrow \mathcal{M}$, 
$(s,k) \longmapsto \mathcal{E}(s,k) = m'$.
\item $inv:\mathcal{K} \longrightarrow \mathcal{K}$, s.t. $\forall k \in
  \mathcal{K}, \forall (s,m)\in \mathcal{S}\times\mathcal{M},
  \mathcal{E}(\mathcal{I}(s,m,k),inv(k))=m$.
\item $\mathcal{I}(s,m,k)$ and $\mathcal{E}(c,k^\prime)$ can be computed in
  polynomial time. 
\end{itemize}
 $\mathcal{I}$ is called the insertion 
function, $\mathcal{E}$ the extraction function, $s$ the host content,
$m$ the hidden message, $k$ the embedding key, $k'=inv(k)$ the extraction key, and
$s'$ is the stego-content. If $\forall k \in \mathcal{K}, k=inv(k)$, the stegosystem is symmetric, otherwise
it is asymmetric.
\end{definition}

\begin{definition}[Bounded stegosystem]
Let $f:\mathds{N} \rightarrow \mathds{N}$.
 A stegosystem  $(\mathcal{I},\mathcal{E}, inv)$ of  $(\mathcal{S}, \mathcal{M}, \mathcal{K})$ 
is $f-$bounded if $dom(\mathcal{I}) \subseteq  \displaystyle{\bigcup_{n \in \mathds{N}} \mathfrak{A}^n \times \mathfrak{A}^{f(n)}\times \mathcal{K}}$.
\end{definition}

\begin{definition}[Probability set]
A \emph{probability set} $\mathcal{X} = \left\{ (S_n, P_n), n \in \mathds{N} \right\}$ on $\mathfrak{A}$ 
is an infinite family of couples of finite sets $S_n\subseteq \mathfrak{A}^*$ together with their
probability distributions $P_n$, such that for every $n \in \mathds{N}$, there exists $r \in \mathds{N}$
such that each element of $S_n$ is in $\mathfrak{A}^r$. The integer $r$ is 
denoted $\ell(S_n)$. Moreover it is required that $\ell(S_n)$ is bounded by
a polynomial in $n$.
\end{definition}
%

\subsection{Cryptographically secure hiding}

%
%
%
%

\begin{definition}[Secure hiding]
Let $f:\mathds{N}\longrightarrow \mathds{N}$, $(\mathcal{I},\mathcal{E}, inv)$ 
a stegosystem of  $(\mathcal{S}, \mathcal{M}, \mathcal{K})$, and
$\mathcal{X} = \left\{ (S_n, P_n), n \in \mathds{N} \right\}$ a probability set.
$\mathcal{I}$ is a $f-$secure hiding for $\mathcal{X}$ if it is $f-$bounded and if
for every positive polynome $\mathfrak{p}$, for any probabilistic 
polynomial-time algorithm $\mathcal{D}$, for any $k \in \mathcal{K}$,
for all sufficiently large $i$'s, for all $m \in \mathfrak{A}^{f(\ell(S_i))}$,
\begin{equation}
 \left| Pr\left(\mathcal{D}\left(\mathcal{I}(S_i,m,k)\right)=1\right)
 -Pr\left(\mathcal{D}\left(S_i\right)=1\right)\right|<\dfrac{1}{\mathfrak{p}(i)}+\dfrac{1}{\mathfrak{p}(k)}
\end{equation}
\end{definition}
where the probabilities are taken over the distribution $\mathcal{X}$ as well as over the
coin tosses of $\mathcal{D}$ and where $\varepsilon(k)\rightarrow 0$ when
the size of $k$ grows up. 

Intuitively, it means that there is no polynomial-time probabilistic algorithm
being able to distinguish the host contents from the stego-contents

\begin{proposition}
If $f_1\leq f_2$ one can compute from any $f_2-$secure hiding stegosystem
for $\mathcal{X}$ a 
$f_2-$secure hiding stegosystem for $\mathcal{X}$.
\end{proposition}

\begin{example}
Assume that $\mathfrak{A}=\{0,1\}$, and that 
\begin{itemize}
\item there exists $\alpha\in \mathbb{N}$ such that for every $n$,
  $|S_n|=2^{\ell(S_n)-\alpha}$ and,
\item for each $n$, there exists $s_n\in \{0,1\}^*$ such that 
$S_n=\{s_n\cdot w\mid  w\in  \{0,1\}^{\ell(S_n)-\alpha}$, where $\cdot$ is
  the concatenation product and,
\item for each $n$, each $s\in S_n$, $P_n(s)=1/|S_n|$.
\end{itemize}

Let $G$ be a cryptographic PRNG. We consider that $\mathcal{I}(s,m,k)$ is
defined for $s\in S_n$ and $m\in \{0,1\}^{\ell(S_n)-\alpha}$, for $k$ such
that the length of $G(k)$ is $\ell(S_n)-\alpha$, by $I(s,m,k)=s_n\cdot
(m\oplus G(k))$. Let $f_0$ be the function mapping $n$ into $n/\alpha$. The
symetric stegosystem defined by $\mathcal{I}$ and where
$\mathcal{E}(s_nw,k)=w\oplus G(k)$ is a $f_0-$secure hiding stegosystem for
$\mathcal{X}$ .
\end{example}










\section{Conclusion}

We thus intend to propose to these Journ\`ees Codes et St\'eganographie 2012 this
new rigorous approach for steganalysis based on the complexity theory. It is 
inspired by the definitions of security that can usually be found in other
branches of cryptology. We have proposed a notion of \emph{secure hiding}
and presented a first secure hiding scheme. Our intention is to prove this
result during the presentation, and to give further developments, by adapting
such a notion to define using complexity what is a robust watermarking, and what is a message
that cannot be extracted.




\bibliographystyle{plain}
\bibliography{mabase}

\begin{thebibliography}{10}

\bibitem{BarniBF03}
Mauro Barni, Franco Bartolini, and Teddy Furon.
\newblock A general framework for robust watermarking security.
\newblock {\em Signal Processing}, 83(10):2069--2084, 2003.
\newblock Special issue on Security of Data Hiding Technologies, invited paper.

\bibitem{Cachin2004}
Christian Cachin.
\newblock An information-theoretic model for steganography.
\newblock In {\em Information Hiding}, volume 1525 of {\em Lecture Notes in
  Computer Science}, pages 306--318. Springer Berlin / Heidelberg, 1998.

\bibitem{Cayre2008}
F.~Cayre and P.~Bas.
\newblock Kerckhoffs-based embedding security classes for woa data hiding.
\newblock {\em IEEE Transactions on Information Forensics and Security},
  3(1):1--15, 2008.

\bibitem{Cayre2005}
F.~Cayre, C.~Fontaine, and T.~Furon.
\newblock Watermarking security: theory and practice.
\newblock {\em IEEE Transactions on Signal Processing}, 53(10):3976--3987,
  2005.

\bibitem{Furon2002}
T.~Furon.
\newblock Security analysis, 2002.
\newblock European Project IST-1999-10987 CERTIMARK, Deliverable D.5.5.

\bibitem{Kalker2001}
T.~Kalker.
\newblock Considerations on watermarking security.
\newblock pages 201--206, 2001.

\bibitem{Ker06}
Andrew~D. Ker.
\newblock Batch steganography and pooled steganalysis.
\newblock In Jan Camenisch, Christian~S. Collberg, Neil~F. Johnson, and Phil
  Sallee, editors, {\em Information Hiding}, volume 4437 of {\em Lecture Notes
  in Computer Science}, pages 265--281, Alexandria, VA, USA, July 2006.
  Springer.

\bibitem{Mittelholzer99}
Thomas Mittelholzer.
\newblock An information-theoretic approach to steganography and watermarking.
\newblock In Andreas Pfitzmann, editor, {\em Information Hiding}, volume 1768
  of {\em Lecture Notes in Computer Science}, pages 1--16, Dresden, Germany,
  September 29 - October 1. 1999. Springer.

\bibitem{Perez06}
Luis Perez-Freire, F.~Pérez-gonzalez, and Pedro Comesaña.
\newblock Secret dither estimation in lattice-quantization data hiding: A
  set-membership approach.
\newblock In Edward~J. Delp and Ping~W. Wong, editors, {\em Security,
  Steganography, and Watermarking of Multimedia Contents}, San Jose,
  California, USA, January 2006. SPIE.

\bibitem{Shannon49}
Claude~E. Shannon.
\newblock Communication theory of secrecy systems.
\newblock {\em Bell Systems Technical Journal}, 28:656--715, 1949.

\bibitem{Simmons83}
Gustavus~J. Simmons.
\newblock The prisoners' problem and the subliminal channel.
\newblock In {\em Advances in Cryptology, Proc. CRYPTO'83}, pages 51--67, 1984.

\end{thebibliography}

\end{document}